\documentclass[conference]{IEEEtran}
\usepackage{cite}
\usepackage{amsmath,amssymb,amsfonts}
\usepackage{algorithmic}
\usepackage{graphicx}
\usepackage{textcomp}
\usepackage{xcolor}
\usepackage[hyphens]{url}
\usepackage{caption}
\usepackage{pifont}
\usepackage{makecell}
\usepackage{titlesec}
\usepackage{booktabs}
\usepackage{threeparttable}
\usepackage{arydshln}
\usepackage{multirow}
\usepackage{subfigure}
\renewcommand\arraystretch{1.3}
\def\BibTeX{{\rm B\kern-.05em{\sc i\kern-.025em b}\kern-.08em
    T\kern-.1667em\lower.7ex\hbox{E}\kern-.125emX}}

\title{MES-Attacks: Software-Controlled Covert Channels based on Mutual Exclusion and Synchronization}

\author{\normalsize{Chaoqun Shen, Jiliang Zhang, Gang Qu}}

\author{\IEEEauthorblockN{Chaoqun Shen}
\IEEEauthorblockA{\textit{Hunan University} \\
Changsha, China \\
shencq@hnu.edu.cn}
\and
\IEEEauthorblockN{Jiliang Zhang\footnotemark{*}}
\IEEEauthorblockA{\textit{Hunan University} \\
Changsha, China \\
zhangjiliang@hnu.edu.cn}
\and
\IEEEauthorblockN{Gang Qu}
\IEEEauthorblockA{\textit{University of Maryland, College Park}\\
MD, United States \\
gangqu@umd.edu}
}

\begin{document}
\maketitle
\thispagestyle{plain}
\pagestyle{plain}

\renewcommand{\thefootnote}{\fnsymbol{footnote}}
\renewcommand{\footnoterule}{%
	\kern -3pt
	\hrule width \linewidth
	\kern 2pt 
}
\footnotetext[1]{Corresponding author}

\begin{abstract}

Multi-process concurrency is effective in improving program efficiency and maximizing CPU utilization. The correct execution of concurrency is ensured by the mutual exclusion and synchronization mechanism (MESM) that manages the shared hardware and software resources. We propose {\it MES-Attacks}, a new set of
software-controlled covert channel attacks based on MESM to transmit confidential information. 
MES-Attacks offer several advantages: 1) the
covert channels are constructed at software level and can be deployed on any hardware; 2) closed share of resource ensures the quality of the channels with low interference and makes them hard to be detected; and 3) it utilizes the system's software resources which are abound and hence difficult to isolate. We built covert channels using different MESMs on Windows and Linux, including \texttt{Event}, \texttt{Timer}, \texttt{FileLockEX}, \texttt{Mutex}, \texttt{Semaphore} and \texttt{flock}. 

Experimental results demonstrate that these covert channels can achieve transmission rate of 13.105 kb/s, 12.383 kb/s, and 6.552 kb/s, respectively in the scenarios of local, cross-sandbox and cross-VM, where the bit error rates are all under 1\%.

\end{abstract}

\section{Introduction}
Modern processors share hardware and software resources, such as caches, branch predictors and library codes, among multiple cores, processes, and threads for performance enhancement and resource optimization. However, such resources sharing could bring serious security issues. An attacker can steal secret information via covert channels by manipulating and perceiving the state changes of shared resources\cite{Arcangeli09, Suzaki11}, or performs 
transient execution attacks to disclose sensitive data \cite{Meltdown2018,Spectre2019,Esmaei2018,Foreshadow2018}. 
In this paper, we consider the problem of how a malicious process, which has access to secret data but cannot communicate with other processes, can leak the secret data, and propose MES-Attacks, a novel set of covert channel attacks based on the synchronization and mutual exclusion of shared resources.  

Covert channels can be classified as hardware resource-based channels and software resource-based channels. In the former, hardware resources such as caches \cite{FF2016,FR2014,PP2015,Streamline21}, memory \cite{DRAMA16}, branch predictors \cite{Jumpover16,Dmitry15,Dmitry16} and buses \cite{WuXW15} are utilized to create covert channels. This requires reverse engineering of the hardware resources \cite{DRAMA16,Jumpover16,PaccagnellaLF21}, which is normally expensive, and 
can be alleviated effectively by isolation techniques \cite{BRB19,ShiSCZ11,ZhouRZ16} because it is hardware specific. Software resource-based covert channels \cite{PCA19,GaoSGKPW21}, on the other hand, overcome these drawbacks and are drawing more attentions recently. However, one challenge for the existing software resource-based channels to become practical is how to reduce the bit error rate (BER) of the transmitted data because these channels are vulnerable to interference. This is a direct result of the fact that they use resources that are accessible to all users, which we refer to as {\it open shared resources}, and interference occurs when other processes use or request the resource.

We propose a set of new covert channels based on {\it closed shared software resources} 
that are shared only between the malicious processes, called {\it Trojan} and {\it Spy}, which use the covert channel to send and receive information, respectively.
These covert channels exploit a basic process management mechanism in the operating system (OS), known as the mutual exclusion mechanism and synchronization (MESM) \cite{Vogler00}, which is designed to ensure the correct concurrent execution of multi-processes. 
MESM coordinates the resource usage mong multiple processes. We found it will unintentionally expose the states of processes and hence can be utilized to build covert channels.

More specifically, MES-Attacks transmit information as follows: during synchronization, one process manipulates the execution of other waiting process by the time it satisfies the completion of a condition.
The waiting process infers the transmitted data by measuring the time to release the waiting state. 
During mutual exclusion, one process manipulates another competing process's execution by controlling the time to occupy the resource, and the competing process deduces the transmission data by measuring the time to acquire the resource. 
In the experiments, we build covert channels using different MESMs on Windows and Linux, including \texttt{Event}, \texttt{Timer}, \texttt{FileLockEX}, \texttt{Mutex}, \texttt{Semaphore} and \texttt{flock}. Experimental results demonstrate that 
these covert channels can achieve transmission rate (TR) and BER of 13.105 kb/s, 0.654\%; 12.383 kb/s, 0.683\%; 6.552 kb/s, 0.713\% respectively in the scenarios of local, cross-sandbox and cross-virtual machine (VM). 

The main contributions of this work are summarized as follows.
\begin{itemize}
\item We investigated the classic MESM for 
its feasibility of building covert channels between the Trojan and Spy and implemented several such channels in real 
multi-processes concurrent operating systems.
\item We built the first cooperation-based covert channel. Compared to existing channels, it simplifies the preparation of attacks, reduces the impact on the system, and makes it difficult to be detected.
\item Both mutual exclusion-based and synchronization-based covert channels we constructed are at software level and can be deployed on any hardware, and they utilize the OS's software resources which are abundant and hence difficult to isolate.
\item We carried out rigorous experiments to demonstrate that, through the proposed covert channels, information could be leaked with high TR and low BER in real scenarios of local, cross-sandbox and cross-VM.
\end{itemize}

The remaining of the paper is organized as follows. Section 2 gives background information on covert channels, synchronization, and mutual exclusion. Section 3 describes the threat model. Section 4 analyses the underlying principles of how MESM-based channels can be constructed, describes an overview of the attacks and details the MES-Attacks proposed for the different MESMs. Section 5 introduces the experimental setup, presents the results in three cases, and analyzes the performance. Section 6 discusses the impact of multi-symbol coding on TR.
Section 7 describes related work. Finally, we make a conclusion in Section 8.

\begin{figure}
 	\setlength{\abovecaptionskip}{0.2cm}
 	\centering
 	\includegraphics[width=85 mm, height=50 mm]{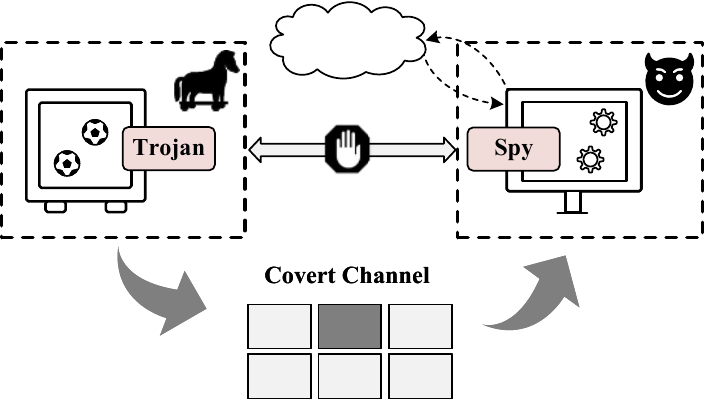}
 	\caption{The model of covert channels.}
 	\label{1}
 	\vspace{-10 pt}
\end{figure}

\section{Background}\label{Pre and Rel}

In this section we briefly discuss the working principle and classification of covert channels, and then introduce the concept and characteristics of MESMs. 

\subsection{Covert Channels}

Covert channel includes two malicious process: Trojan and Spy. Commonly, Trojan can access secret data, but cannot communicate directly with other processes. In contrast, Spy can access overt channels but not secret data \cite{Powertchannel19}. The covert channel is established on the collaboration of such two processes, as shown in Fig.\ref{1} . Concretely, the Trojan process (sender) induces the system resource changes, which would be observed by the Spy process (receiver) to infer information such as, cryptography key \cite{YaoDV18}.

\begin{figure}[t]
	\setlength{\abovecaptionskip}{0.2cm}
 	\centering
 	\includegraphics[width=0.95\linewidth]{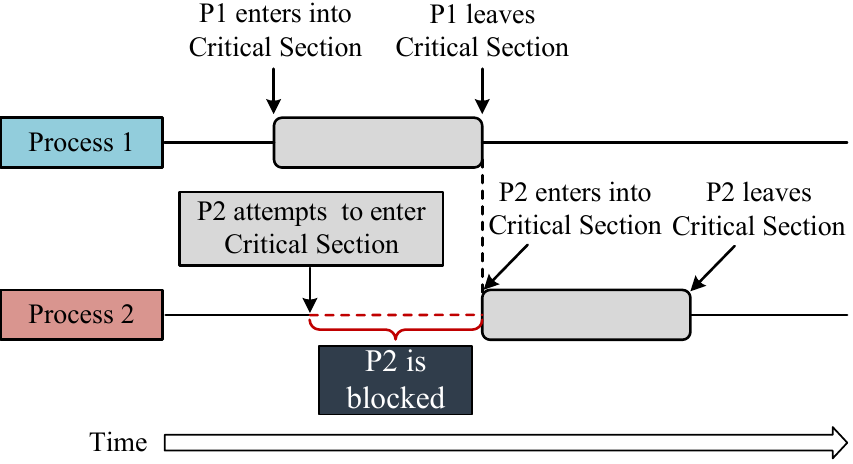}
 	\caption{The schematic representation of process mutual exclusion.}
 	\label{3}
 	\vspace{-15 pt}
\end{figure}

Covert channels can be categorized into volatile channels and persistent channels, depending on whether the state of the altered resource is retained on the system \cite{XiongS21}. In the volatile channel, the Trojan and the spy run concurrently and both dynamically compete for hardware resources. On the basis of the secret data to be sent, the Trojan decides whether to cause contention on a shared resource, while the spy measures changes to the shared resource in real time. In the case of the port covert channel, for example, when sending a `1', the Trojan occupies the port. At the same time, the Spy continuously utilizes the same port and deduces the secret data via observing the time change in real time to infer the port's occupancy. In the persistent channel, the Trojan and the Spy can execute asynchronously and the state of the shared resource is retained for a period of time. Based on the secret data to be sent, the Trojan decides to place the shared resource in a different state, while the spy asynchronously measures the change in the shared resource. In the cache covert channel, for example, when sending `1', the Trojan loads the data into cache. Without concurrent execution, the spy waits for a pre-calculated period of time and then observes the time change to infer the cache state change and thus deduce the secret data.

\subsection{Process Mutual Exclusion}
Process mutual exclusion is indirect constraints between processes. 
When a process enters the critical section to use a critical resource, another process that needs to use the same resource must wait. Only when the process currently occupying the critical resource exits the critical section, then the waiting process can be released from the blocking state, as shown in Fig. \ref{3}. The critical resource mentioned here refers to a resource that only one process is allowed to use at a time. And a critical section is a program segment in a process that accesses a critical resource.
For example, process 2 haves to access a variable, but process 1 is making use of the same variable at the moment. Then process 2 will be blocked until process 1 has released that resource.


\begin{figure}
 	\setlength{\abovecaptionskip}{0.2cm}
 	\centering
 	\includegraphics[width=75 mm, height=58 mm]{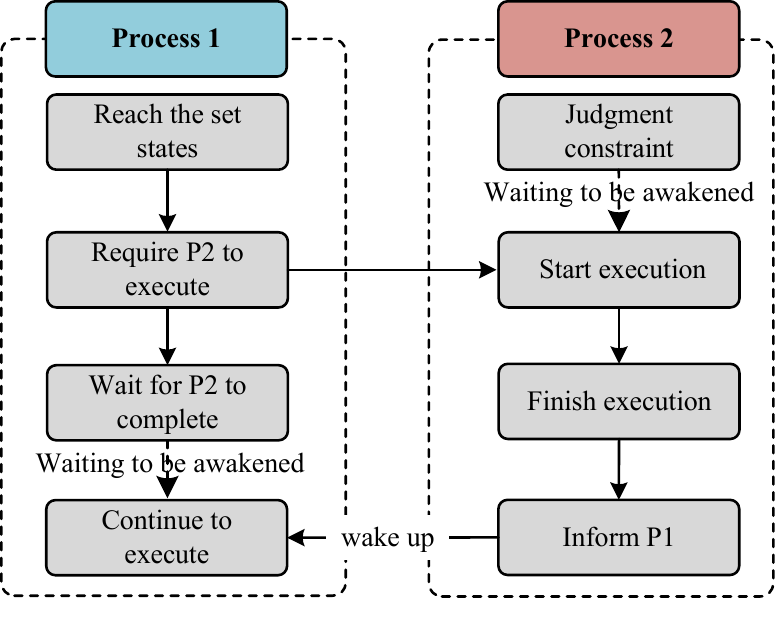}
 	\caption{The schematic representation of process synchronization.}
 	\label{2}
   	\vspace{-12 pt}
\end{figure}

\subsection{Process Synchronization}

Process synchronization is a direct constraint relationship between processes that refers to multiple processes waiting or passing information to coordinate the order of their work at certain points, as shown in Fig.\ref{2}. The moment to pause and wait is called the synchronization point; the operation to release the wait or the information sent by other processes is called the synchronisation condition.
For example, process 1 must read the information generated by process 2 from a buffer, and when the buffer is empty, process 1 will be blocked. Only when the target data is placed in the buffer, it will be woken up.


For ease of illustration, we refer to the mutual exclusion and synchronization states collectively as the constraint states in the following.

\section{Threat Model}

We adopt the common covert channel model \cite{PCA19,GaoSGKPW21,Powertchannel19}, where a Trojan process has collected secret data from a restricted environment but cannot communicate with other processes directly, how can the Trojan process transmit the data outside the restricted environment? More specifically, we have implemented the following three scenarios: {\it local} where the Trojan wants to send information to another process on the same machine, {\it cross-sandbox} where the Trojan is inside a sandbox and wants to send data out of the sandbox, and {\it cross-VM} where the Trojan wants to send data from one VM to another. 

We consider the most general cases and do not require the attacker to possess any specific capabilities. 
More specifically, we only assume that 
in the local and cross-VM scenarios, we assume that Trojans and spies can select shared resources by prior agreement, but have no write ability and only read permission;
while in sandbox scenarios, processes in the sandbox cannot write directly to external resources due to the security policy of the sandbox itself. 
In summary, in all scenarios, there is a need to ensure that both processes cannot write directly to shared resources.

Questions such as how the Trojan gets into the restricted environment, how it obtains sensitive data, and what is the size of the data are out of the scope of this paper. We focus on how can the attacker establish a covert channel such that the Trojan process (sender) can transmit the data out to a Spy process (receiver).

\section{The Proposed MES-Attacks}
 
In this section, we discuss the construction of MESM covert channels. It is worth noting that the MESMs of Linux OS are all inter-thread operations, except for \texttt{flock}, which can be used directly between processes. In order to apply the inter-thread mechanism to processes, the two threads need to be connected to shared memory. However, this shared memory must be set to be readable and writable, defying the principle of covert channel, so we can only use the \texttt{flock} to carry out the attack. For the same reason, Windows OS can build channels using \texttt{Event}, \texttt{Timer}, \texttt{FileLockEX}, \texttt{Mutex}, \texttt{Semaphore} (these are known as kernel objects). 

This section explains why we chose MESMs to build the channel, analyses the underlying principles of MESM channels, followed by an overview of synchronization-based and mutual exclusion-based channels, and finally a detailed discussion of channel design and communication protocols, using \texttt{Event} (synchronization), \texttt{flock} (mutual exclusion) and \texttt{Semaphore} (special mutual exclusion) as examples.

\subsection{Why Choose Process MESMs}

Process communication is divided into low-level communication and high-level communication, depending on the amount of information exchanged and how efficient it is. Low-level communication can only pass state and control information and includes {\it MESM}, {\it signal} etc. High-level communication improves the efficiency of signal communication, enables the transfer of large amounts of data and reduces programming complexity, including {\it shared memory}, {\it pipe}, {\it message}, etc.

It is clear from this that high-level communication methods are used directly for the transfer of specific data information. The signals sent by MESMs between processes can only be used for the purpose of coordinating the execution of processes by informing them whether to continue waiting or to move on. These signals themselves do not contain any data information.

However, with MESMs, the time at which a process lifts its constraint state is directly controlled by the other processes. We make the signals sent by MESM have data information by encoding the time at which a process releases its constraint state (see Section IV.C for details). Meanwhile, MESMs are common and easily controlled in all major OS. Therefore, we construct covert channels to transmit secret information through the design of MESM executions. Note that other low-level communication methods such as {\it signal} may also be able to be used to design covert channels, and this is left for our future work.

\subsection{Communication Principle of MES-Attacks}

\subsubsection{How channels work on Windows}

\begin{figure}
 	\setlength{\abovecaptionskip}{0.2cm}
 	\centering
 	\includegraphics[width=0.995\linewidth]{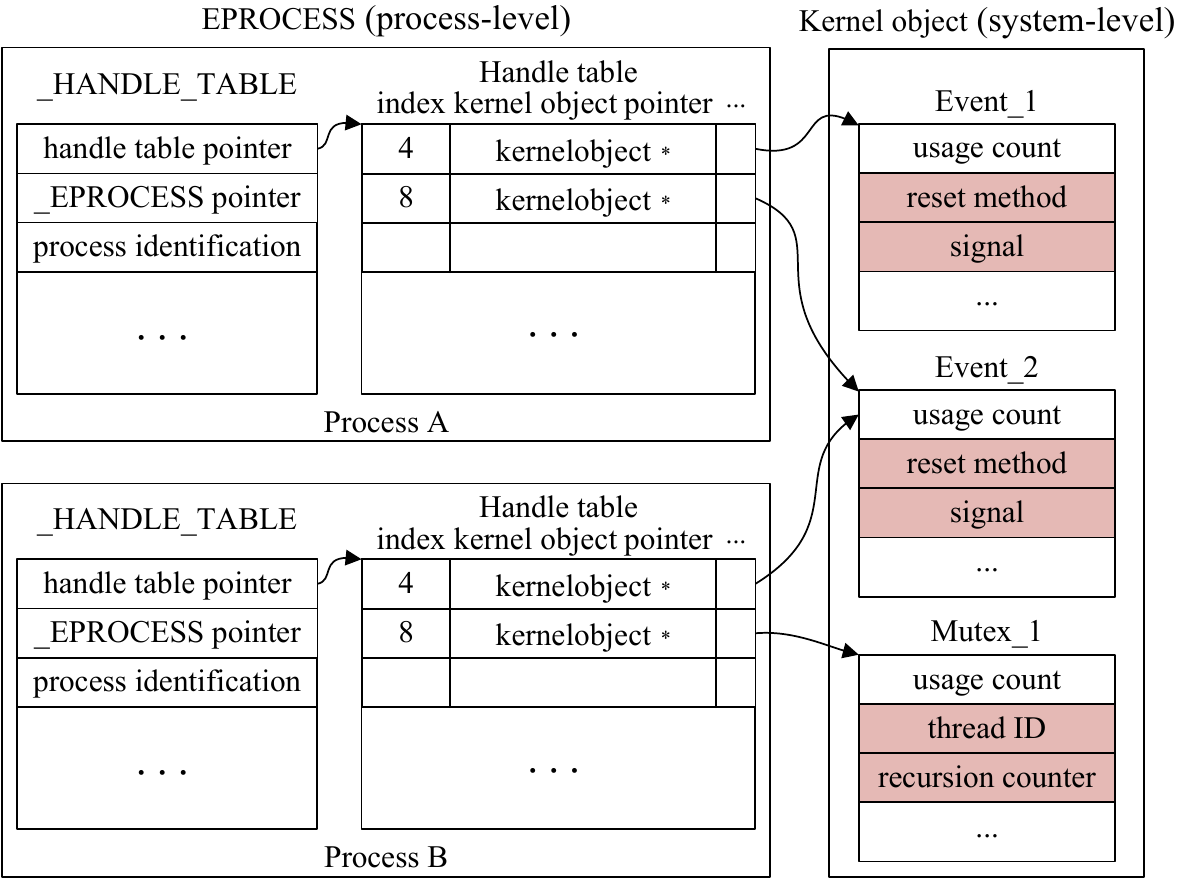}
 	\caption{The underlying principle of channels in Windows.}
 	\label{4-1}
  	\vspace{-12 pt}
\end{figure}

We make use of kernel objects under Windows, namely \texttt{Event}, \texttt{Mutex}, \texttt{Semaphore} and \texttt{WaitableTimer}.
Kernel objects are data structures maintained by the operating system and are the basic interface between user-mode code and kernel-mode code.
As the kernel object is at the OS level, different processes can access the same kernel object. To avoid direct modification of the kernel object by user-level programs, processes access it via a created handle. Each process has its own handle table, in which only the handles to kernel objects are stored. Handles that point to the same kernel object do not have the same value among different processes, and handles with the same handle value usually point to different kernel objects. The details are shown in Fig.\ref{4-1}.

The kernel object needs to work with WaitForSingleObject to implement the synchronization mutual exclusion function.
The purpose of WaitForSingleObject is to wait for a specified object until it is in a signalled state or beyond a set interval of time.
In this experiment we have set the wait time to infinity so that the execution of the WaitForSingleObject is determined only by the signalling state of the object.
The signal state here is a data member of the kernel object, where ``signalled'' means that the object is available to the thread, and ``unsignalled'' means that the thread is waiting for the object.
\texttt{Event} also makes use of the data member `reset', which indicates the Event is automatically or manually set to the no-signal state after a successful call to WaitForSingleObject.
In the case of the \texttt{Mutex}, the signalling state of the object is characterised by the thread ID and the recursion counter.
The data members of other different kernel objects also differ, and the execution of WaitForSingleObject will vary.

As shown in Fig.\ref{4-1}, process A creates the kernel object Event\_2, which is automatically reset and has an initial state of no signal, and waits for a signal change in Event\_2 via WaitForSingleObject. Process B opens Event\_2 via handle 8, and when some operation is completed, the process sets Event\_2 to the signalled state. At this point, process A's WaitForSingleObject ends its wait and continues down the execution. From the above analysis, we can observe that the OS achieves inter-process synchronization by the fact that the object can be requested or not. Therefore, we can construct a covert channel by modulating the time when a Spy process pointing to the same kernel object as the Trojan process is able to request the object.

\begin{figure}
 	\setlength{\abovecaptionskip}{0.2cm}
 	\centering
 	\includegraphics[width=\linewidth]{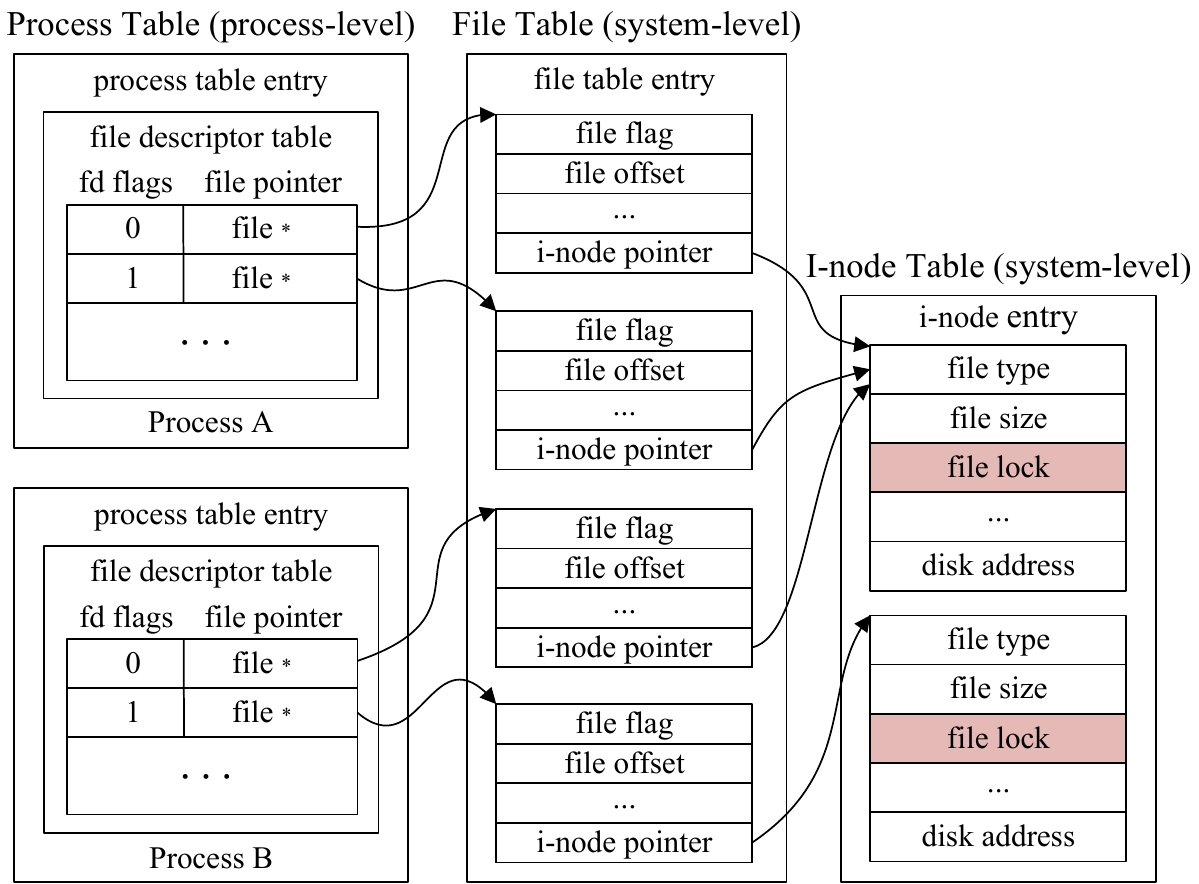}
 	\caption{The underlying principle of channel in Linux.}
 	\label{4-2}
 	\vspace{-12 pt}
\end{figure}

\subsubsection{How channels work on Linux}

We make use of file locks in Linux.
Every file that is opened in the kernel needs to be maintained by three data structures, namely the file descriptor table, the file table and the i-node table.
The relationship between them is shown in Fig.\ref{4-2}.
Each time a process is started, the system allocates a process table entry to it, which contains the file descriptor table.
The file descriptor table stores all the file descriptors (fd) opened by the process and pointers to the corresponding file table entries. This is a process-level data structure, which means that each process has its own file descriptor table.
The kernel also maintains a file table which stores information about files opened by all processes in the system.
This is a system-level data structure that can be shared between processes.
Each open file then corresponds to an i-node data structure,
which stores real file information such as file type, size, owner, access permissions, file's address table.
i-node table is also a system-level data structure and is not unique to the process.

It is worth noting that when the same file is opened by either the same or different processes, independent file table entries corresponding to each fd are created.
And these fds share information about the file. 
Therefore, \texttt{flock} enables process mutual exclusion, explained in terms of the relationship between processes and files in Fig.\ref{4-2}. When process A uses \texttt{flock} to put an exclusive lock on fd1, the locking information is added to the i-node table entry.
Since the OS can detect that process A's fd1 and process B's fd0 point to the same i-node, then processes A and B share the file corresponding to the i-node. 
At this point process B will fail to add any further type of file lock to fd0 and the OS will block it.
Based on the above analysis, we design the covert channel by blending the time that processes pointing to the same i-node can lock the file.

\subsection{Overview of MES-Attacks}


\begin{table*}
  \caption{The category of MES-Attacks.}
  \begin{center}
  \label{tab1}
  \begin{tabular}{cccc}
    \toprule
    {\textbf{Mechanism}} & {\textbf{Type}} & {\textbf{Preparatory Work}} & {\textbf{Measurement}}\\
    \midrule
    Synchronization & Cooperation & Setting conditions & The time to meet the condition \\
    Mutual exclusion & Contention & Selected critical resources & The time to lock\\
    \bottomrule
  \end{tabular}
  \end{center}
 \vspace{-10 pt}
\end{table*}

\begin{figure*}
 	\setlength{\abovecaptionskip}{0.2cm}
 	\centering
 	\includegraphics[width=0.9\linewidth]{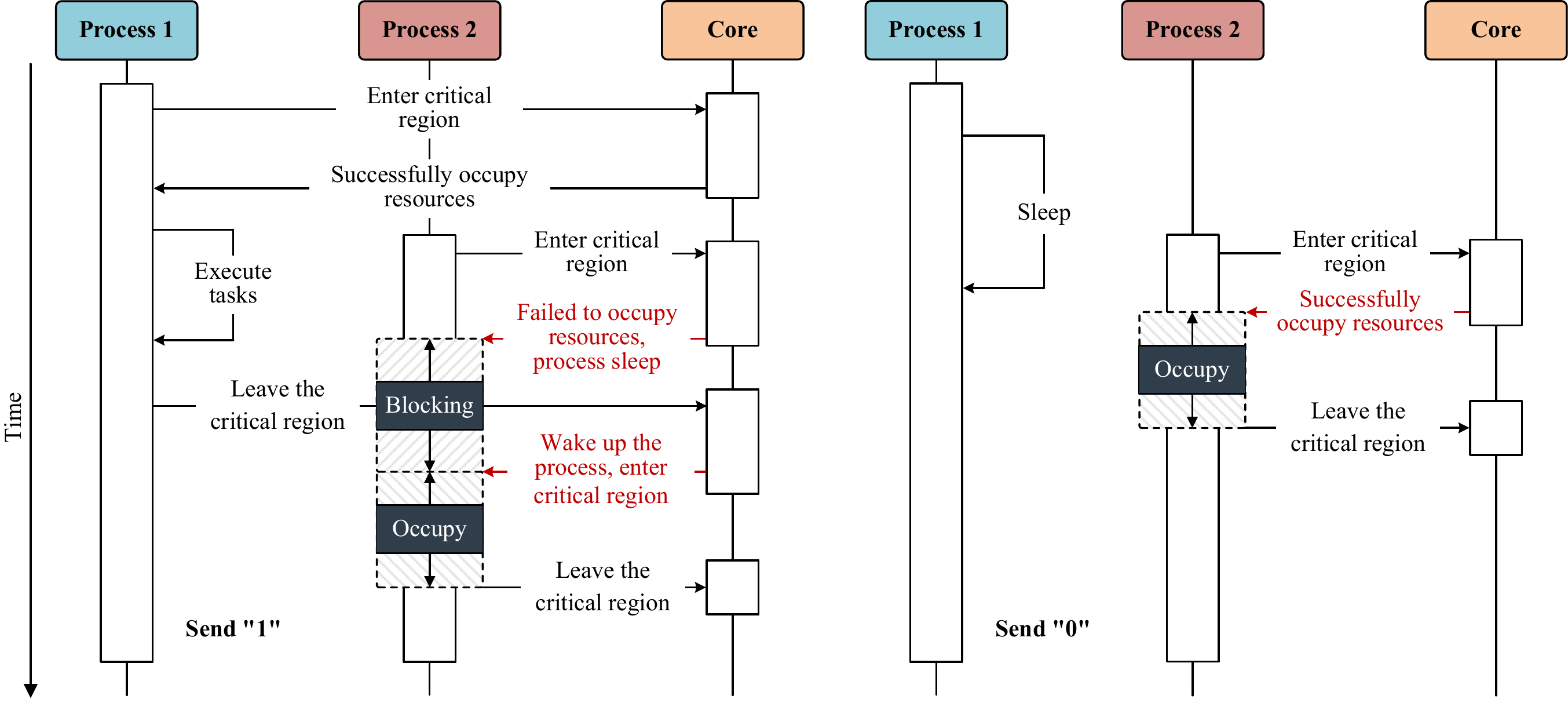}
 	\caption{The working principle of mutual exclusion mechanism-based covert channels.}
 	\label{6}
\end{figure*}

\begin{figure}
 	\setlength{\abovecaptionskip}{0.2cm}
 	\centering
 	\includegraphics[width=0.9\linewidth]{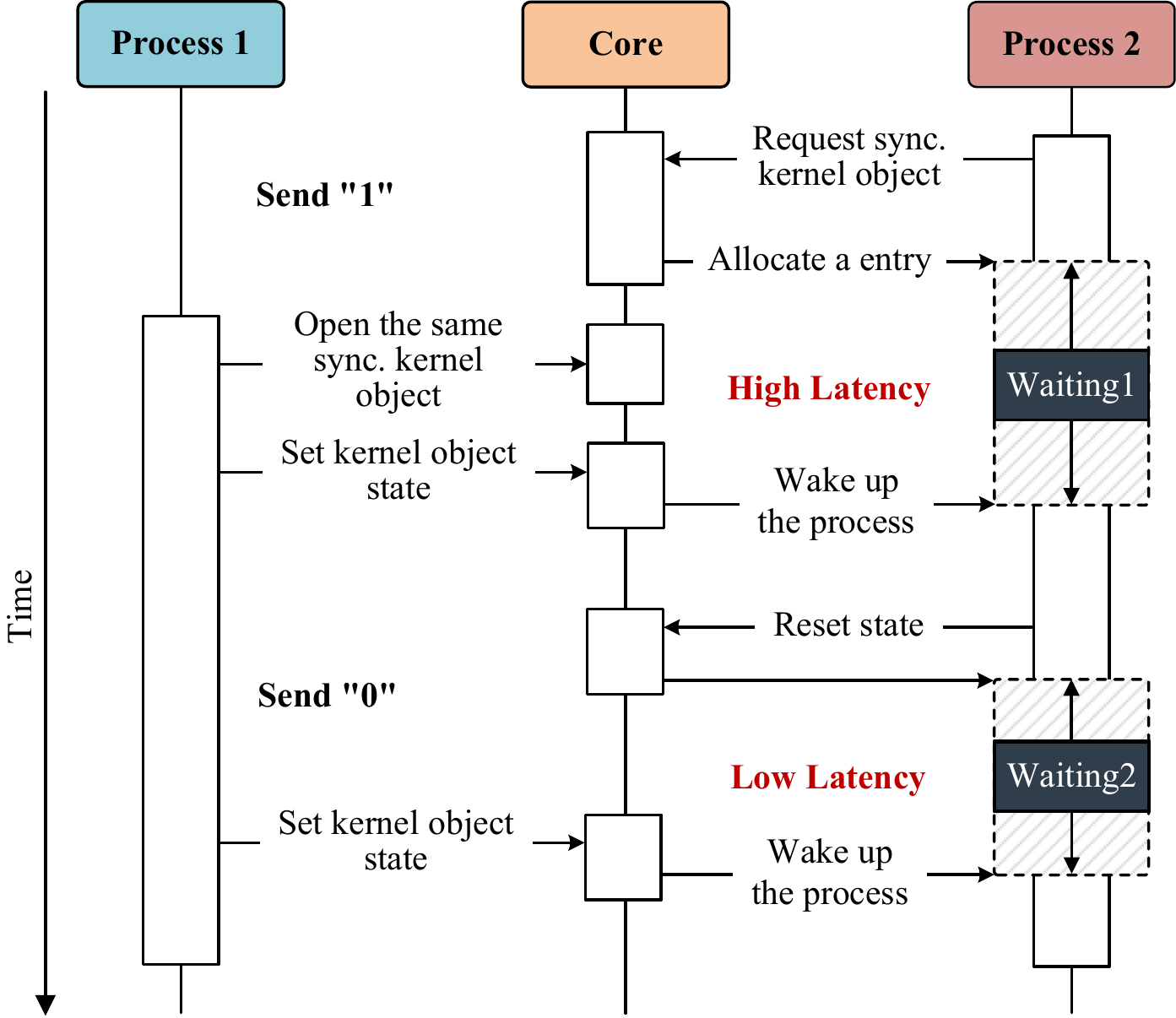}
 	\caption{The working principle of synchronization mechanism-based covert channels.}
 	\label{5}
 	\vspace{-12 pt}
\end{figure}

In the proposed MES-Attacks, the Trojan and the Spy must perform concurrently. Importantly, the secret data is recovered by the spy through the real-time monitoring of changes in its own execution time, and the recovery is based on changes in resources rather than the state of the retained resources. Therefore, the covert channels we have designed are volatile covert channels.

Existing volatile covert channels are all implemented based on inter-process contention, but we have found something different. Mutual exclusion is a guarantee that critical resources can only be used by one process at the same time, then under the mutual exclusion mechanism, Trojans and Spies are regulating time through a competitive relationship, in line with the characteristics of existing volatile channels. However, the situation is different under the synchronization mechanism, where Trojans and Spies are based on a cooperative relationship to leak data. In this case, the Spy can only continue execution after the Trojan has met a certain condition. In the process of communication, they do not compete with each other for shared resources, but rather cooperate with each other to satisfy the execution condition. Thus, the synchronization-based channel we implement is a novel type of covert channel --- a cooperation-based volatile covert channel. The category of MES-Attacks as shown in Table \ref{tab1}.

\begin{table}[ht]
    \centering
    \renewcommand{\arraystretch}{1.1}
    \begin{tabular}{l}
    \hline
    \textbf{Protocol 1} \texttt{flock}-based covert channels.\\
    \hline
    \textbf{sender's key code ()} \{\\
    1: \quad char key[$n$]; \\
    2: \quad fp = open (``file.txt'', ``r+b''); // Open the shared file\\
    3: \quad for (int $i$ = 0; $i$ $\textless$ $n$; $i$++) \{\\
    4: \quad \quad if (key[$i$] $==$ 1) \{\\
    5: \quad \quad \quad flock (fp --\textgreater fileno, LOCK\_EX);\\
    6: \quad \quad \quad sleep (RESTRICTION\_PERIOD);\\
    7: \quad \quad \quad flock (fp --\textgreater fileno, LOCK\_UN); \}\\
    8: \quad \quad sleep (SLEEP\_PERIOD\_1); \} \}\\

    \hdashline[1 pt/1 pt]
    \\
    \hdashline[1 pt/1 pt]
    \textbf {receiver’s key code ()} \{\\
    1: \quad fp = open (``file.txt'', ``r+b'');\\
    2: \quad while(1) \{\\
    3: \quad \quad start\_time;\\
    4: \quad \quad flock (fp --\textgreater fileno, LOCK\_EX);\\
    5: \quad \quad flock (fp --\textgreater fileno, LOCK\_UN);\\
    6: \quad \quad end\_time;\\
    7: \quad \quad if (end\_time $-$ start\_time $\textgreater$ threshold)\\
    8: \quad \quad \quad key[$i++$] = 1;\\
    9: \quad \quad else\\
    10:\quad \quad \quad key[$i++$] = 0;\\
    11:\quad \quad sleep (SLEEP\_PERIOD\_2);\\
    12:\quad \quad if (i $\geq$ n)\\
    13:\quad \quad \quad break; \}\}\\
    \hline
    \end{tabular}
    \label{Protocol1}
\end{table}

The Trojan can encode data through regulating the execution time of the Spy.
Concretely, when encoding `1', the Trojan enters a mutually exclusive or synchronous state and takes up resources for longer time.
When encoding `0', in the mutual exclusion case, the Trojan enters sleep state directly for shorter time. In the case of synchronization,  the Trojan still enters the synchronization state, it only takes up resources for shorter time than sending `1'.
Thus, the Spy can clearly distinguish the `1' from the `0' by measuring the time to exit the constraint state, \textit{i.e.}, high latency decoding for logical `1' and low latency decoding for logical `0'.
Fig.\ref{6} shows the working principle of contention-based channels. The Trojan (Process 1) controls the blocking time of the Spy, and the Spy (Process 2) can infer the data sent by the Trojan based on the time of relieving mutual exclusion state is $t$ = $t_{blocking}$ + $t_{occupy}$ or $t = t_{occupy}$. Fig.\ref{5} shows the working principle of cooperation-based channels. The Trojan (Process 1) changes its time to wake up the Spy, and the Spy (Process 2) can infer the data sent by the Trojan with the time of relieving synchronization state is $t$ = $t_{waiting1}$ or $t$ = $t_{waiting2}$. 

It is worth noting that \cite{noteinweb} explores the problem of program execution constraints and mentions, in general terms, six different situations that can cause information leakage. Point 5 mentions that ``If the system has interlocks which prevent files from being open for writing and reading at the same time, the service can leak data if it is merely allowed to read files which can be written by its owner'', followed by a brief description of the 1 bit transfer.

However, it is flawed: firstly, the fundamental problem is that the file property exploited by \cite{noteinweb} is readable and writable, then the attacker can transmit information by directly reading and writing to that file, without the need to use it to construct a covert channel. Secondly, the paper does not show the communication protocol, and the transfer example given is rather concise, while the synchronization process is cumbersome and seriously affects performance. Finally, as \cite{noteinweb} is only a high level illustration of information leakage by the author, no experiments and analysis are performed, the performance of the channel is not measured. It is not a complete technical work. In our work, we set the file attribute to read-only and exploit the mandatory locking of the file to prohibit an attacker from transmitting information directly through the file. Further,  we analyse in detail the problem of communication synchronization and verify the performance of the covert channel in multiple scenarios on multiple systems.

\begin{table}[ht]
    \centering
    \renewcommand{\arraystretch}{1.1}
    \begin{tabular}{l}
    \hline
    \textbf{Protocol 2} \texttt{Event}-based covert channels.\\
    \hline
    \textbf{sender's key code ()} \{\\
    1: \quad char key[n]; \\
    2: \quad HANDLE handle = OpenEvent (trojan\_event); \\
    \quad \quad // Open the event object created by the Trojan\\
    3: \quad for (int $i$ $=$ 0; $i$ $\textless$ $n$; $i$++) \{\\
    4:\quad \quad if (key[$i$] $==$ 1) \{ \\
    5:\quad \quad \quad sleep (RESTRICTION\_PERIOD\_1);\\
    6:\quad \quad \quad SetEvent (handle);\}\\
    7:\quad \quad else \{ \\
    8:\quad \quad \quad sleep (RESTRICTION\_PERIOD\_2);\\
    9:\quad \quad \quad SetEvent (handle);\} \} \}\\
    \hdashline[1 pt/1 pt]
    \\
    \hdashline[1 pt/1 pt]
    \textbf{receiver's key code ()} \{\\
    1: \quad HANDLE handle = CreatEvent (trojan\_event);\\
    2: \quad while(1) \{\\
    3: \quad \quad start\_time;\\
    4: \quad \quad WaitForSingleObject (handle, INFINITE);\\
    5: \quad \quad end\_time;\\
    6: \quad \quad  if (end\_time $-$ start\_time $\textgreater$ threshold)\\
    7: \quad \quad \quad key[$i++$] $=$ 1;\\
    8: \quad \quad  else\\
    9: \quad \quad \quad key[$i++$] $=$ 0;\\
    10:\quad \quad if (i $\geq$ n)\\
    11:\quad \quad \quad break;\} \}\\
    \hline
    \end{tabular}
    \label{Protocol2}
\end{table}

More importantly, we explore the MESMs, a process management mechanism in OS, of which lock is only one of the contention-based channels, and we propose three other contention-based channels and innovatively propose cooperation-based channel.


\subsection{Covert Channel based on \texttt{flock}}

A file lock guarantees that only one process can access one file at any given time. This mechanism makes reading and writing files more secure. In Linux, \texttt{flock} is described as an example, which uses LOCK\_EX and LOCK\_UN to build a covert channel. Other lock functions that implement locking, unlocking and blocking, such as read locks, can also be used to build covert channels and are not discussed here.


The relevant communication protocol for \texttt{flock} is \textbf{Protocol 1}. When encoding `1', the Trojan enters mutual exclusion state; otherwise, the Trojan enters sleep state. In order to maintain transmission synchronization, SLEEP\_PERIOD\_1 = SLEEP\_PERIOD\_2. The communication protocols of other covert channels based on contention are similar.

\subsection{Covert Channel based on \texttt{Semaphore}}

The \texttt{Semaphore} is equivalent to an identifier of resource which is used for processes to coordinate the utilization of the resource such as recording the number of resources, waiting processes or blocking queues. 
The \texttt{Semaphore} is implemented by P/V atomic operations (OS primitives). We set a \texttt{Semaphore} $S$ representing the number of resources. The P operation is to decrease the $S$ by 1, and V operation does the opposite. After the process performs the P/V operation, if $S$ $\textgreater$ 0, the process can obtain resources and continue to execute, if $S$ $\textless$ 0, the process will be blocked.


\begin{table*}[htbp]
\begin{minipage}[t]{0.48\textwidth}
\caption{Unprocessed implementation for \texttt{semaphore}}
\makeatletter\def\@captype{table}
\begin{center}
\setlength{\tabcolsep}{3.5mm}
\begin{tabular}{cccc}
\hline
\textbf{Key} & \textbf{Trojan}   & \textbf{Spy} & \textbf{Resources}\\
\hline
K1 & Request & Release   &  0 \\
K2 & Request & Release   &  \textbf{\textcolor{red}{0}} \\
K3 & \textbf{\textcolor{red}{Sleep}} & Unable to release   &  0 \\
K4 & Request & Release   &  0 \\
K5 & Request & Release   &  \textbf{\textcolor{red}{0}} \\
K6 & \textbf{\textcolor{red}{Sleep}} & Unable to release   &  0 \\
K7 & Request & Release   &  \textbf{\textcolor{red}{0}} \\
K8 & \textbf{\textcolor{red}{Sleep}} & Unable to release   &  \textbf{\textcolor{red}{0}} \\
K9 & \textbf{\textcolor{red}{Sleep}} & Unable to release   &  \textbf{\textcolor{red}{0}} \\
K10 & \textbf{\textcolor{red}{Sleep}} & Unable to release   &  0 \\
K11 & Request & Release   &  0 \\
K12 & Request & Release   &  0 \\
\hline
\end{tabular}
\begin{tablenotes}
    \footnotesize
      \item[1] \textbf{K} = 1,1,0,1,1,0,1,0,0,0,1,1. 
      \item[2] \textbf{Initial Resources} = 0.
\end{tablenotes}
\label{tab2}
\end{center}
\end{minipage}
\begin{minipage}[t]{0.48\textwidth}
\caption{Improved implementation for \texttt{semaphore}}
\makeatletter\def\@captype{table}
\begin{center}
\setlength{\tabcolsep}{5mm}
\begin{tabular}{cccc}
\hline
\textbf{Key} & \textbf{Trojan}   & \textbf{Spy} & \textbf{Resources}\\
\hline
K1 & Request & Release   &  5 \\
K2 & Request & Release   &  5 \\
K3 & Sleep & Release   &  4 \\
K4 & Request & Release   &  4 \\
K5 & Request & Release   &  4 \\
K6 & Sleep & Release   &  3 \\
K7 & Request & Release   &  3 \\
K8 & Sleep & Release   &  2 \\
K9 & Sleep & Release   &  1 \\
K10 & Sleep & Release   &  0 \\
K11 & Request & Release   &  0 \\
K12 & Request & Release   &  0 \\
\hline
\end{tabular}
\begin{tablenotes}
    \footnotesize
      \item[1] \textbf{K} = 1,1,0,1,1,0,1,0,0,0,1,1. 
      \item[2] \textbf{Initial Resources} = 5.
\end{tablenotes}
\end{center}
\label{tab3}
\end{minipage}
\vspace{-12pt}
\end{table*}

In a \texttt{Semaphore}-based covert channel, the Trojan decides whether to produce resources based on the data bits sent, and Spy infers the data based on the time of resource released. Unlike previous MESMs, the channels have additional preparation operations: the ``shared buffer'' need to be set between two processes, and there are a certain number of initial resources. Otherwise, when the Trojan sends 0, the Spy will come to a standstill as it cannot complete the operation to release the resources. As shown in Table \ref{tab2}, assuming that the data to be sent $K$ = \{1,1,0,1,1,1,0,1,0,0,0,1,1\}, the number of initial resources in $S$ is 0. The first column of the table refers to the $n$-th bit of the transmission $K$. The fourth column indicates the remaining number of resources in $S$ after sending one bit data. The second and fourth columns, marked in red, indicate that when there are no resources in $S$ and the next bit to be transferred is `0', the Trojan does not produce resources and the Spy stalls. For example, the ``Resources'' of $K2$ is 0, and $K3$ goes to sleep state without producing resources. At this time, the Spy is blocked without output until the next time the Trojan producing resources (sends `1'). Therefore, the final output of the Spy is determined by the number of `1' in the key $K$, making it impossible to recover the data correctly. In order to ensure that each transmitted bit has a correct time measurement, the attack needs to prepare a certain amount of resources in $S$ ahead of time, and $S$ should be at least equal to the number of `0' in the confidential data. As shown in Table \ref{tab3}, the initial number of resources is 5.

Of all the MESMs we have utilised, only the \texttt{Semaphore} itself can be used for both synchronization and mutual exclusion. However, it cannot take advantage of the ``generate resources before consuming them'' principle of working, as this would result in the spy receiving no `0'. Therefore, the Trojan and Spy cannot be executed in a specific order, that is, the synchronization function of \texttt{Semaphore} cannot be used.


\subsection{Covert Channel based on \texttt{Event}}

Due to the uncertainty of the process execution sequence, errors may occur. Therefore, to ensure that processes are executed in the correct order, a process that does not reach the setting condition should block itself. The OS achieves this by means of a synchronization mechanism.

\begin{figure}
 \setlength{\abovecaptionskip}{0.2cm}
 \centering	\includegraphics[width=88 mm, height= 34mm]{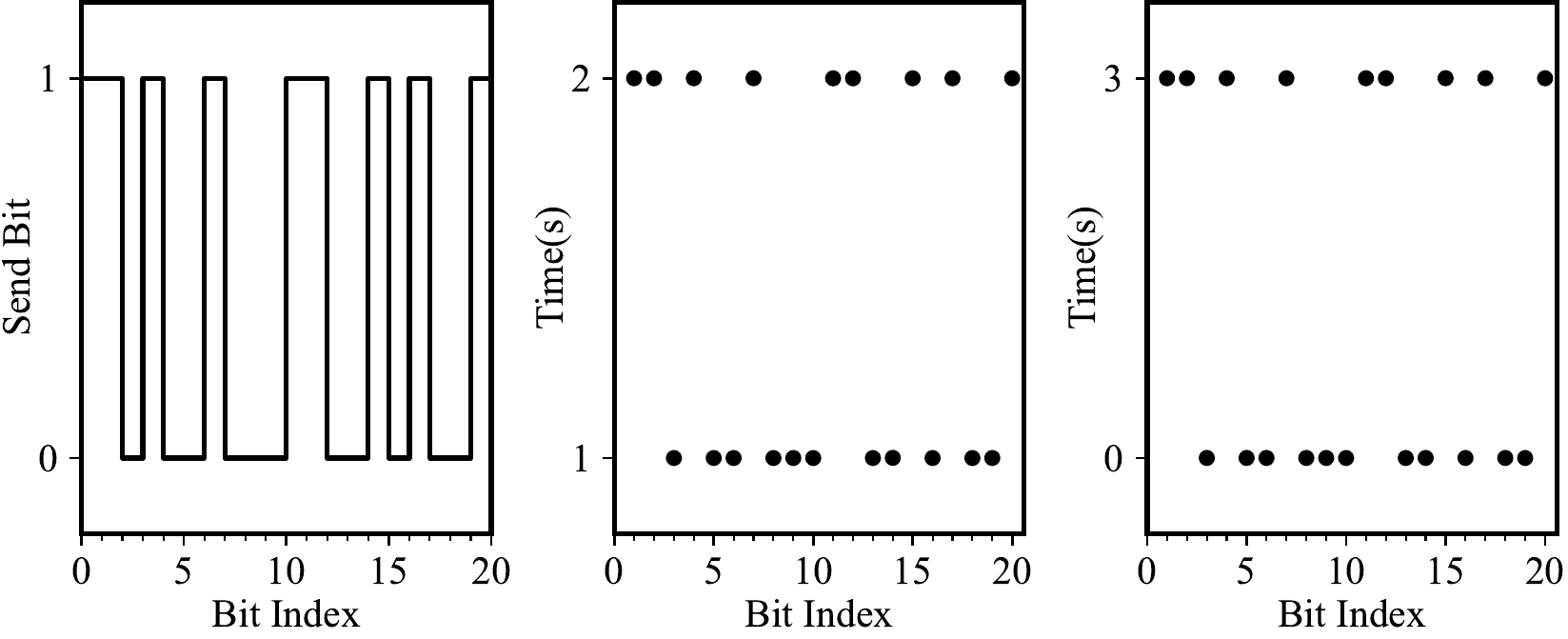}
 	\caption{The proof of concept for MES-Attacks. (a) data sent by the Trojan, (b) the Spy under synchronization, and (c) the Spy under mutual exclusion.}
 	\label{7}
 	\vspace{-15pt}
\end{figure}

Take Event as an example, it indicates whether the process can occupy the handle by its ``signalled'' and ``unsignalled'' states. To achieve inter-process synchronization, these objects are used in a wait function --- WaitForSingleObject. The wait function allows a thread to block its own execution until the specified unsignalled object is set to the signalled state. Concretely, the Trojan process calls the SetEvent function to set the state to ``signalled'', which wakes up the Spy process's WaitForSingleObject, thus releasing the Spy process from its blocking state and continuing its execution.

The communication protocol for \texttt{Event} is shown in \textbf{Protocol 2}. The sender enters the synchronization state when it sends a `0' or `1'. However, the constraint time of sending the `0' and `1' is different, \textit{i.e.}, RESTRICTION\_PERIOD\_1 $\neq$ RESTRICTION\_PERIOD\_2. The communication protocols of other covert channels based on cooperation are similar.

\subsection{Summary}

The proposed MES-Attacks utilizes mutual exclusion and synchronization mechanisms to achieve contention-based and cooperation-based covert channels, respectively.
Under the mutual exclusion mechanism, we select \texttt{flock}, \texttt{FileLockEx}, \texttt{Mutex} and \texttt{Semaphore} to construct contention-based covert channels. Then, under the synchronization mechanism,  we select \texttt{Event} and \texttt{Timer} to construct cooperation-based covert channels. 
It should be noted that not all exploitable MESMs are listed in this paper and only the six cases mentioned above are used as examples.

As shown in Fig.\ref{7}, we conducted a proof-of-concept to demonstrate the feasibility of MES-Attacks. Fig.\ref{7}(a) shows that the transmission sequence is \{1,1,0,1,0,0,1,0,0,0,1,1,0,0,1,0,1,0,0,1\}. In  Fig.\ref{7}(b), when sending `1', the Trojan needs to wait for 2s to meet the agreed condition. Otherwise, wait for 1s. In Fig.\ref{7}(c), when sending `1', the Spy lock shared resources for 3s. Otherwise, the Spy enters the sleep state for 1s. Fig.\ref{7}(b) and (c) depict the detection time of the Spy. The experimental results show that `1' and `0' are distinguishable.

Compared with previous covert channels, it has significant advantages as follows:

\textbf{\ding{192} Closed resources}. The previous covert channels exploit open shared resources, which is accessible to all processes and hence introduces considerable interference during the communication between the Trojan and Spy. For example, in a cache-based covert channel, the data is loaded or not loaded, which is determined not only by the Trojan but also by other processes. In contrast, in MES-Attacks, the software resources being exploited are shared objects or files specified by Trojan and Spy negotiations in advance. This closed resources setup make it difficult for other processes to exhaust and access. As a result, the interference of other processes in the attack can be greatly reduced.

\textbf{\ding{193} Cooperative channel}. It has several advantages: a) Simpler preparation. First, the persistent covert channels must perform additional state recovery operations at every bit-period between the Trojan and Spy. For example, in cache-based channels, the Trojan needs to perform operations such as flush, eviction and cache-thrashing for each transmission bit before the next bit. On the contrary, the cooperation-based channel has no residual state and can therefore  communicate continuously according to the agreed number of bits. 
Secondly, the existing covert channels require a long synchronization sequence to ensure correctness, while the cooperation-based covert channel only requires a very short synchronization sequence (e.g., one bit);
b) Smaller BER. In the previous covert channels, once a bit error occurs, a series of subsequent bits would be affected. However, in the cooperation-based channel, since the end of the spy's measurement is determined only by the time when the Trojan arrives at the synchronization point. This causes the measurement status of the next bit of data to be the same after sending one bit of data. Therefore, only one bit error occurs, i.e., the previous bit error does not affect the subsequent bit correctness (bit independence); 
c) Higher system efficiency. Most of covert channels consume resource frequently. For example, in a port-based covert channel, the sender and the receiver need to take turns occupying port resources, thereby reducing the system efficiency. While the cooperative covert channel uses only one shared variable and controls the transmission bits through sleep time during communication, which reduces the utilization of system resources and improves efficiency.
%


It is remarkable that MES-Attacks are universal and powerful convert channel attacks because the inter-process MESMs are basic functions in modern OSs.

\begin{figure*}
\centering
\setlength{\abovecaptionskip}{-0.1cm}
\subfigure[The relationship between time and BER.]{
\centering
\includegraphics[width=0.46\linewidth]{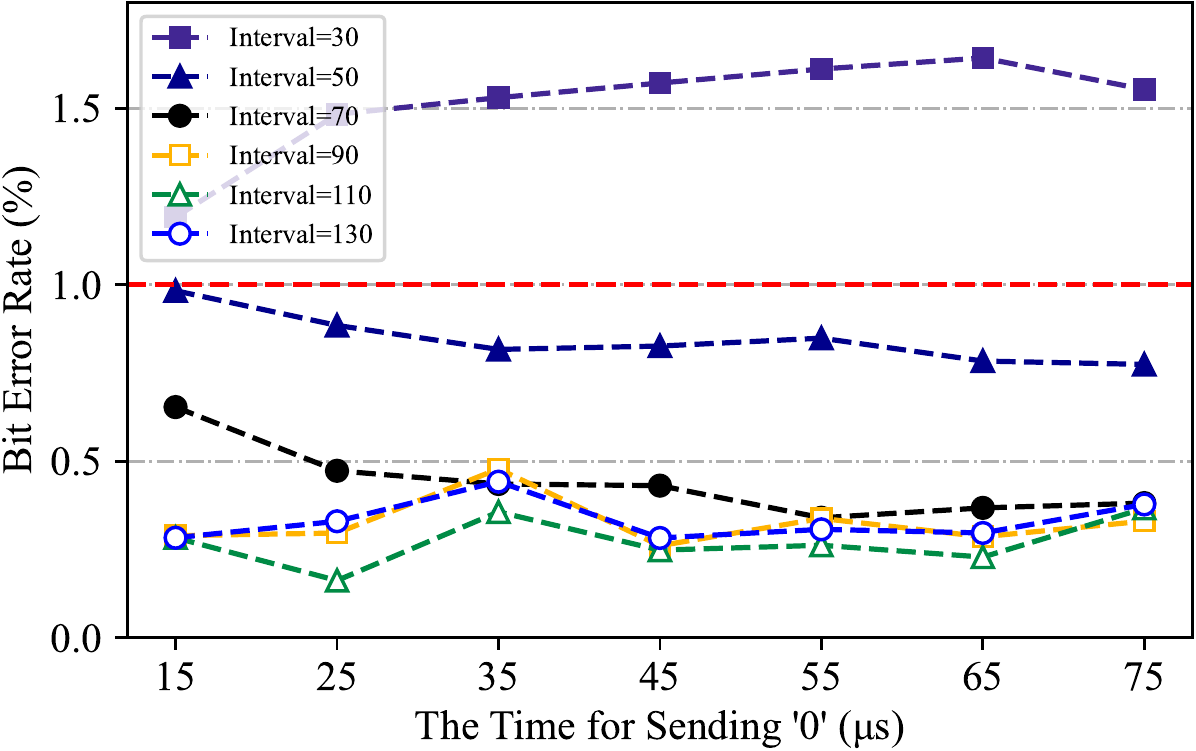}}
\quad
\subfigure[The relationship between time and TR.]{
\centering
\includegraphics[width=0.46\linewidth]{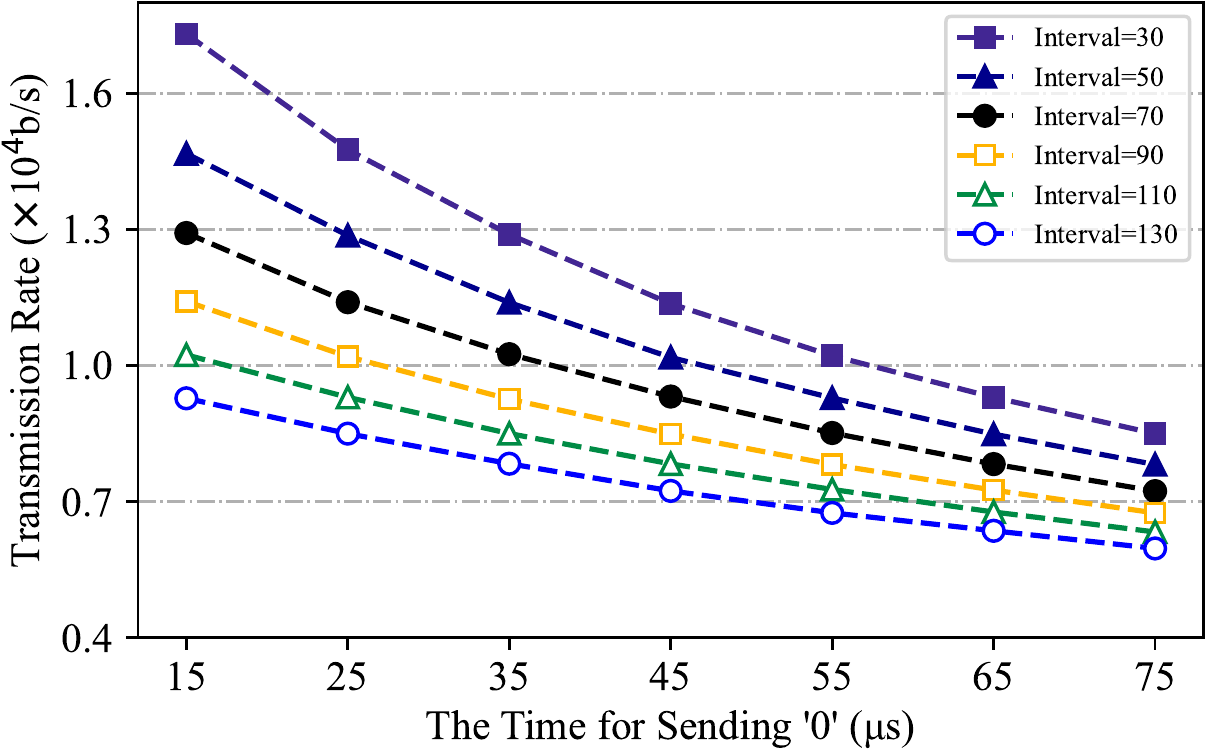}}
\caption{The performance impact of timing parameters on \texttt{Event}-based covert channels.}
\label{8}
\vspace{-10 pt}
\end{figure*}

\begin{figure}
 	\setlength{\abovecaptionskip}{0.2cm}
 	\centering
 	\includegraphics[width=\linewidth]{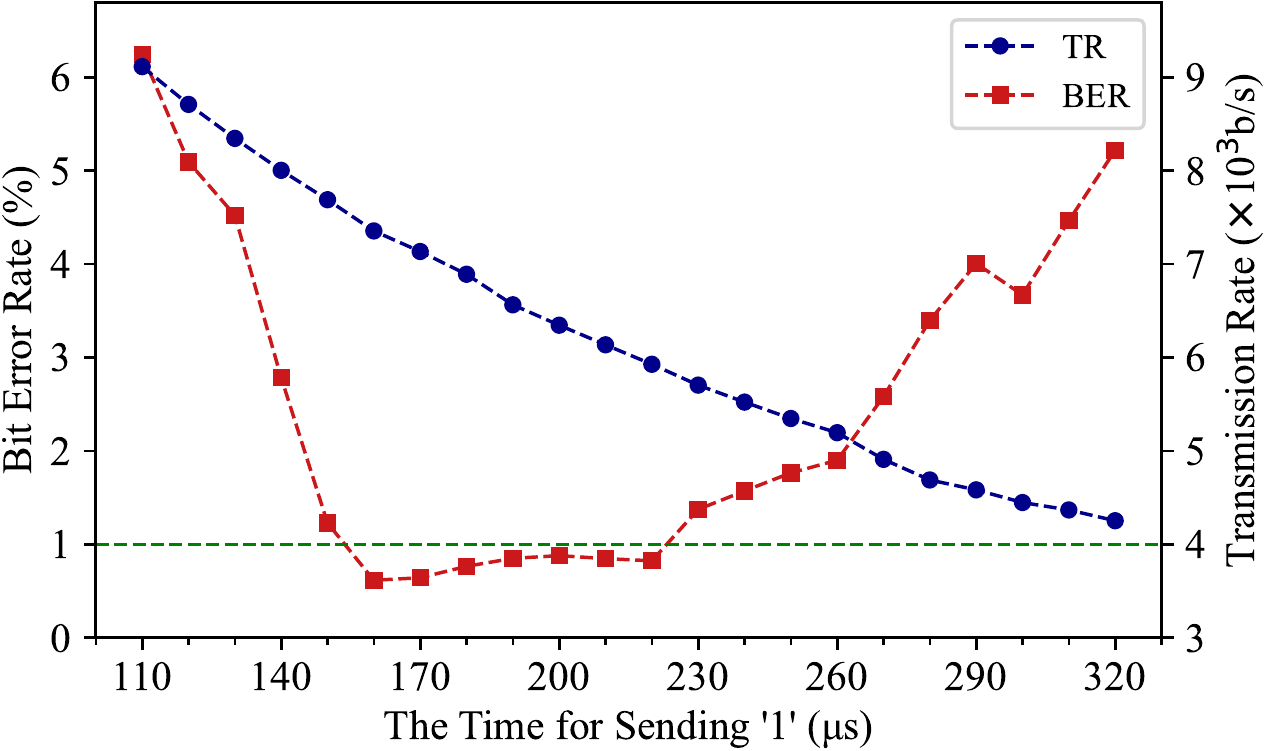}
 	\caption{The performance impact of timing parameters on \texttt{flock}-based covert channels.}
 	\label{9}
 	\vspace{-12 pt}
\end{figure}

\section{Evaluation}

In this section we first illustrate the experimental setup and the implementation of communication synchronization. Then we implement the MES-Attacks and measure the performance in different scenarios on Windows and Linux systems. Finally, we discuss the TR improvement due to symbolic encoding.

\subsection{Experimental Setup}


We conduct experiments on an Intel i5-7400 with 4 cores. The operating system is Ubuntu 16.04 LTS with Linux kernel 4.15.0 and Windows 10. We built covert channels in three scenarios: local, cross-sandbox, and cross-VM.

\subsection{The synchronization of communications}
To ensure that the Spy process successfully receive the secret data sent by Trojan in covert channels, we implement synchronization between the Trojan and Spy at the communication beginning. Note that in this subsection, we discuss synchronization for communication rather than process management. Concretely, the Trojan sends an $n$-bit pre-negotiated bitstream (such as ``10101010''), called ``synchronization sequence'', before transmitting an $m$-bit secret data. The Spy measures the time taken to release the constraint state to infer `0' or `1' and verifies whether the first $n$-bit of the received data are equal to the ``synchronization sequence''. Only when they are equal, the Spy considers the subsequent $m$-bit as secret data. Otherwise, the Spy discards the received data and prepares for the next round. At the end of the communication, a long delay is used as a synchronization signal fed to the Trojan when the Spy received such $n+m$ bits, which stands for this round is over and a new one can launch.

Besides, we also requires a fine-grained synchronization for the transmission of each bit because of the following reasons:


\textbf{\ding{192} The order in which critical resources are occupied affects data transmission.} Each critical resource only supports a process at a time in the mutual exclusion mechanism, which forces such two processes working with competition. However, the mutual exclusion mechanism includes the fair and unfair competes and MES-Attacks can only run in the fair pattern. To be more specific, the fair competition allocates the critical resource according to the queue of process. During the unfair competition, the current process has the privilege to get the resource again with ignoring the process queue. In this case, once the Trojan turns into the sleep state (send out a bit `0'), the critical resource would be occupied by the Spy until the end of current round of transmission, which results in Spy receiving the rest of the bits as 0.

\textbf{\ding{193}  The execution time of each instruction affects the accuracy of the measurement.} During the transmission of data, each instruction that is not related to the constraint state (referred to as ``irrelevant instruction'') takes a certain amount of time to execute. 
Therefore, the execution time $t_{t}$ of the irrelevant instruction between two data sent by the Trojan and the execution time $t_{s}$ of the irrelevant instruction between two data received by the Spy affects the Spy's judgement. If $t_{t}$ $\textgreater$ $t_{s}$, the Spy's measurement time will take the execution time of the Trojan's irrelevant instruction into account; if $t_{t}$ $\textless$ $t_{s}$, the Spy's measurement time will miss part of the Trojan's constraint state time. In addition, such errors are accumulated under the mutual exclusion mechanism.

Hence, a fine-grained inter-bit synchronization between the Trojan and the Spy is required after each bit is transmitted. This ensures that the Trojan and the Spy meet the execution sequence requirements, breaks the Spy's continuous occupation of resources and increases the accuracy of data recovery.


\subsection{Performance}

The time parameter determines the channel's TR and BER. Intuitively, the smaller the time parameter, the better the performance of the channel. However, in practice, if the time parameter is too small, it can exacerbate the effects of factors such as resource scheduling or interrupt handling in the system. Therefore, in this section we carefully discuss the relationship between time parameters and channel performance in local, cross-sandbox and cross-VM scenarios.

\subsubsection{Local}

There are two time parameters in the synchronization-based covert channel: $t_{w0}$, the wait time for sending `0'; $t_{i}$, the interval between sending `0' and sending `1'.

Taking \texttt{Event} as an example, experimental results are shown in Fig.\ref{8}. Fig.\ref{8}(a) shows the relationship between time and BER. 
When $t_{w0}$ $\textless$ 15 \textmu s, the error rate exceeds 1\%. It is difficult for the Spy to capture the `0' due to  the small $t_{w0}$, resulting in a larger error rate. 
$t_{i}$ determines the resolution of the Spy. When $t_{i}$ is too small, the Spy cannot distinguish `0' and `1', resulting in a higher error rate. 
However, in the case of $t_{i}$ = 30 \textmu s, the larger the $t_{w0}$, the higher the error rate. Since the larger $t_{w0}$ indicates that the Trojan takes longer to send `0', the number of times that the system is blocked will increase. It causes a longer time for the Trojan to meet the condition when sending `0' and misleads the Spy to presume the data to be `1'. 
Instead, when $t_{i}$ $\geq$ 50 \textmu s, $t_{w0}$ will no longer affect the error rate, with BER not exceeding 1\%.
Since $t_{i}$ determines the resolution of Spy, only when the $t_{i}$ makes the gap between sending `1' and sending `0' is large enough, the random error caused by the system no more seriously influences the Spy, and the BER remains stable at a low level. At this point, the Spy can distinguish `1' and `0' even if the system block prolongs the time of sending `0'. 
\begin{table*}[htb]
  \caption{The channel performance in local scenario.}
  \begin{center}
  \label{tab4}
  \begin{tabular}{ccccccc}
    \toprule
{\textbf{Attack methods}}&{\textbf{\texttt{flock}}}&{\textbf{\texttt{FileLockEX}}}&{\textbf{\texttt{Mutex}}}&{\textbf{\texttt{Semaphore}}}&{\textbf{\texttt{Event}}}&
    {\textbf{\texttt{Timer}}}\\
    \midrule
    Timeset(\textmu s) & $t_{t1}$ = 160, $t_{t0}$ = 60 & $t_{t1}$ = 150, $t_{t0}$ = 50 & $t_{t1}$ = 140, $t_{t0}$ = 60 & $t_{t1}$ = 230, $t_{t0}$ = 100 & $t_{w0}$ = 15, $t_{i}$ = 65 & $t_{w0}$ = 15, $t_{i}$ = 75\\
    BER(\%) & 0.615 & 0.758 & 0.759 & 0.741 & 0.554 & 0.600\\
    TR(kb/s) & 7.182 & 7.678 & 7.612 & 4.498 & 13.105 & 11.683 \\
  \bottomrule
\end{tabular}
\end{center}
\end{table*}

\begin{table*}[htb]
  \caption{The channel performance in cross-sandbox scenario.}
  \begin{center}
  \label{tab5}
  \begin{tabular}{ccccccc}
  \toprule
    {\textbf{Attack methods}}&{\textbf{\texttt{flock}}}&{\textbf{\texttt{FileLockEX}}}&{\textbf{\texttt{Mutex}}}&{\textbf{\texttt{Semaphore}}}&{\textbf{\texttt{Event}}}&
    {\textbf{\texttt{Timer}}}\\
    \midrule
    Timeset(\textmu s) & $t_{t1}$ = 170, $t_{t0}$ = 60 & $t_{t1}$ = 170, $t_{t0}$ = 60 & $t_{t1}$ = 150, $t_{t0}$ = 60 & $t_{t1}$ = 240, $t_{t0}$ = 100 & $t_{w0}$ = 15, $t_{i}$ = 70 & $t_{w0}$ = 15, $t_{i}$ = 85\\
    BER(\%) & 0.642 & 0.700 & 0.701 & 0.731 & 0.583 & 0.610\\
    TR(kb/s) & 6.946 & 7.181 & 7.109 & 4.338 & 12.383 & 10.458 \\
  \bottomrule
\end{tabular}
\end{center}
\vspace{-10 pt}
\end{table*}

Fig.\ref{8}(b) shows the relationship between time and TR. With the same $t_{w0}$, a larger $t_{i}$ indicates that it takes longer to transmit `1', so the rate will be smaller. In the case of keeping $t_{i}$ unchanged, the larger $t_{w0}$, the longer the time required to transmit `0' and `1', and hence the lower the rate. Therefore, to keep the error rate below 1\% and get the maximum transmission rate, we choose $t_{w0}$ = 15 \textmu s and $t_{i}$ = 70 \textmu s, at which point the TR of the single transmission is 13.105 kb/s. 
Assuming that in the threat model an attacker can control multiple Trojan and Spy processes at the same time, the transmission rate is further increased. The number of simultaneous transfers that the Trojans and Spies can launch is determined by the number of processes that can run concurrently on the system. In addition to the nuances of the way that the processes specify their minimum working sets, the two major factors that determine the answer on any particular system include the amount of physical memory and the system commit limit. Our tests show that the number of concurrent processes on our system is 6833, so ideally we can achieve transfer rates of tens of Mbps. And, the better the performance of the system, for example, the more cores and memory, the greater the threat of attack.

In a mutual exclusion-based covert channel, we take \texttt{flock} on Linux as an example. Since the process scheduling mechanism requires 58 \textmu s to wake up the sleep function, we directly set $t_{t0}$, i.e., the time to sending `0', as 60 \textmu s. However, this problem does not exist in Windows, where the time to send `0' is set more flexibly. Thus, there is only one time parameter in this example: the time $t_{t1}$ to send `1'.

The experimental results are shown in Fig.\ref{9}. The overall error rate changes with time, which shows a relationship similar to the ``concave function''. When 160 us $\leq$ $t_{t1}$ $\leq$ 220 \textmu s, the BER stabilizes below 1\%; when $t_{t1}$ $\leq$ 160 \textmu s, the BER gradually increases with the decreasing of $t_{t1}$ due to the limited accuracy of the Spy to distinguish data (similar to \texttt{Event}-based channel); when $t_{t1}$ $\geq$ 220 \textmu s, the number of times that the system is blocked will increase and the Spy presumes the detected shorter time as `0' due to system blocking, so the error rate increases. The smaller the $t_{t1}$, the greater the TR, so we recommend $t_{t1}$ = 160 \textmu s with an BER of 0.615\%. With such setting, the TR of a single transmission is 7.182 kb/s. Similarly, assume that an attacker can control multiple Trojan and Spy processes  all at once. And the number of processes available for transmission in the system is determined by the number of file descriptors, which defaults to 1024. Ideally, we can achieve a TR of several Mbps.

The performance of MES-Attacks in the local scenario is summarized in Table \ref{tab4}. In a cooperation-based channel, where processes work in conjunction with each other, the time to send `1' and `0' can be set shorter and there is no need to consider fine-grained communication synchronization.
Hence, the TR of a cooperation-based channel is significantly greater than that of a contention-based channel.
\texttt{Semaphore} requires 6 instructions (P-P-S-sleep-V-V) to perform a lock in a contention-based channel, Whereas \texttt{flock}, \texttt{FlieLockEx} and \texttt{Mutex} only require 3 instructions (lock-sleep-unlock), so the TR of \texttt{semaphore} is lower than in the other cases.

\subsubsection{Cross-sandbox}


Sandboxes are often used to test potentially harmful programs or other malicious code that may have access to sensitive information. However, since the processes running inside the sandbox cannot communicate directly with processes on the system, we can build covert channels to transfer data. On Linux systems, we take advantage of the Firejail sandbox. And on Windows systems, we make use of Sandboxie.

\begin{table}
  \centering
  \caption{The channel performance in cross-VM scenario.}
  \label{tab6}
  \begin{tabular}{ccc}
    \toprule
    {\textbf{Attack methods}}&{\textbf{\texttt{flock}}}&{\textbf{\texttt{FileLockEX}}}\\
    \midrule
    Timeset(\textmu s) & $t_{t1}$ = 200, $t_{t0}$ = 70 & $t_{t1}$ = 190, $t_{t0}$ = 70 \\
    BER(\%) & 0.832 & 0.713 \\
    TR(kb/s) & 5.893 & 6.552\\
  \bottomrule
\end{tabular}
\captionsetup{skip=10pt}
\vspace{-15pt}
\end{table}

Table \ref{tab5} shows the performance of the MES-Attacks in the cross-sandbox scenario. Its changing trend of TR is the same as the local scenario (see Table \ref{tab4}). Compared with the local scenario, the cross-sandbox scenario takes longer time to transmit the data because the Trojan and Spy need to ``break'' the isolation mechanism provided by the sandbox.

\subsubsection{Cross-VM}

According to Section IV.B, Windows kernel objects can be exploited to design covert channels because the information of kernel objects in the system is shared among processes. Our \texttt{FileLockEX} test experiments with Hyper-V and VMware Workstation Pro 15 showed that only in Hyper-V did the two VMs wait for each other. We suspect this is because Hyper-V is a Type I VM and its Virtual Machine Manager (VMM) is installed directly on the hardware, so the kernel objects opened in the OS on it are shared. VMware Workstation Pro 15, on the other hand, is a Type II VM, where the VMM is installed on the host OS, resulting in kernel objects between VMs are not shared. Therefore, we select Hyper-V on Windows systems. \texttt{flock} on Linux is a covert channel via a file table entry, and selecting KVM is sufficient.

Note that there is a problem with the implementation of covert channels in the VM scenario: 
only \texttt{FileLockEX} is valid, the rest of the channels do not work. The reason for the situation is that the kernel objects utilised by the FileLockEX-based channels point to read-only files that are shared between VMs; the other objects created do not correspond to real resources just generate an identity, thus these objects only exist in their respective sessions, i.e. they are isolated between VMs. 
The solution is to find an object that is multiplexed between the VM and the host, and then have the two VMs build a channel through that multiplexed object. Some system service applications satisfy this condition (e.g. wininit.exe), but the names of the objects generated by these applications are hidden. The future work is therefore to find the names of the shared objects.

\begin{figure}
 	\setlength{\abovecaptionskip}{0.2cm}
 	\centering
 	\includegraphics[width=\linewidth]{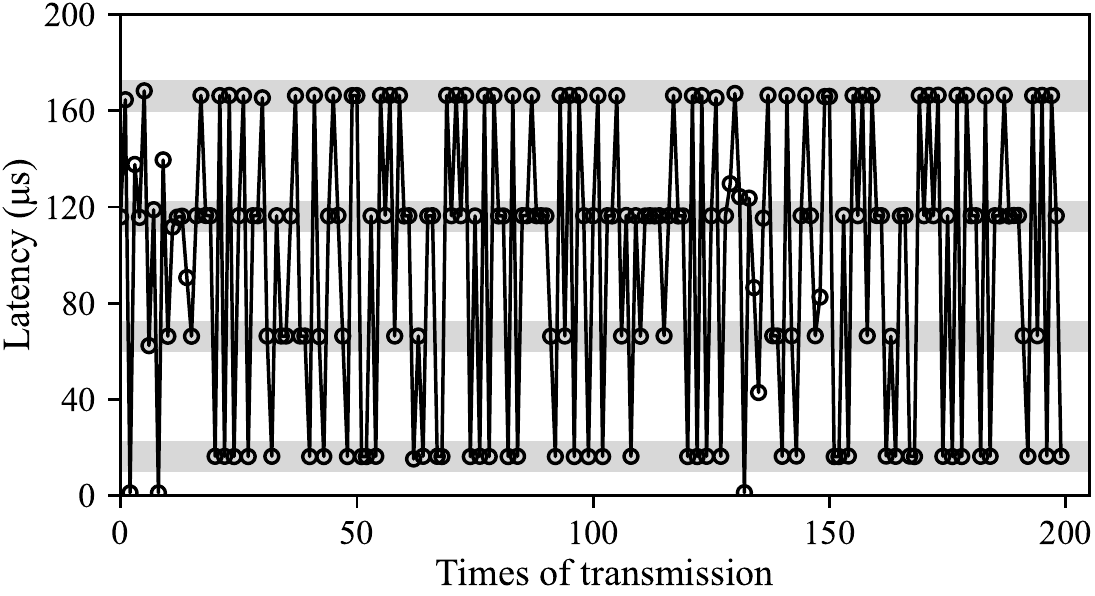}
 	\caption{Mutli-bit symbol tranmission using 4 combination pairs to encode 2-bit symbols.}
 	\label{10}
 	\vspace{-12 pt}
\end{figure}

Table \ref{tab6} shows the performance in a cross-VM scenario. The reason of decreasing in TR is the Trojan and the Spy are on different VMs and the information transfer between them needs to go through paths that become longer.


\section{Symbols Encoding Multi-bits}

The TR of a channel is determined by the number of code elements transmitted per unit time and the amount of information each code element carries. The previous experiments discussed the case of binary code elements where `0' and `1' are transmitted with the same probability. This section explores ways to further increase the TR by using multiple bit encodings to increase the amount of information contained in the transmitted code elements, using the example of an event in a local scenario on Windows.

As can be seen from Fig.\ref{8}(a), the BER within $t_{i}$ $\geq$ 50 \textmu s is acceptable.
For this reason, we combine several different wait times within this time range to encode different signals. 
For example, when we are encoding a 2-bit symbol, the Trojan encodes `00', `01', `10', `11' by setting the time to perform SetEven to 15 \textmu s, 65 \textmu s, 115 \textmu s and 165 \textmu s.
Fig.\ref{10} shows the 2-bit symbol transmission which all four distinct symbols are observed.
We note that encoding 2-bit symbol achieves a higher TR, with experiments showing a peak transmission rate of approximately 15.095 kb/s, higher than the 13.105 kb/s observed for 1-bit symbol transmission.
However there is no further increase in rate when encoding 3-bit (or more) symbols. The reason for this is that there are 8 time setting cases at this point, the number of judgement cases that need to be performed when transmitting data increases and the time required to transmit data such as `101', `110' and `111' becomes larger. In combination, the TR does not increase.

\section{Related Work}

According to the used resource types, covert channels can be classed into hardware resource-based and software resource-based.

\subsection{Hardware resource-based covert channels}

The hardware components shared in the micro-architecture hierarchy can be used to build covert channels, and cache covert channels are the most discussed. Cache and memory have different locations and manufacturing devices in the processor, so there is a time difference between data obtained from cache and memory. Based on this fact, cache side-channel attacks such as Prime + probe \cite{PP2015}, Flush + Reload \cite{FR2014}, Flush + Flush \cite{FF2016},  Evict + Time \cite{GrasRBBG17}, Prime + Abort \cite{PA17}, Prime + Scope \cite{PS21} are proposed. 
Besides, cache mechanisms such as cache consistency protocol \cite{YaoDV18}, cache eviction strategy \cite{LRU20,BriongosMM020}, and cache compression \cite{Safecracker20} can also be exploited. 

In addition to cache side-channel attacks, DRAMA \cite{DRAMA16} can build covert channels by using the time difference between row hit and row conflict. Address translation units like MMU \cite{GrasRBBG17} and TLB \cite{TLB18}, prediction units like PHT \cite{BranchScope18}, BTB \cite{Jumpover16} and way predictor \cite{LippH0PMG20} can also be used.
Due to the limited total amount of resources, another kind of channel leaks data through whether there is resource contention, including memory bus contention \cite{WuXW15}, port contention \cite{AldayaBHGT19,SMoTherSpectre19}, CPU interconnection channel contention \cite{AhnKKJDSJK21,Meshup22} and so on. 

\subsection{Software resource-based covert channels} 
Software resource-based side channel attacks, including page deduplication-based attacks \cite{BosmanRBG16,SuzakiIYA11} and browser cache-based attacks \cite{JiaDLS15,JacksonBBM06a,GoethemJN15,AciicmezSK05}. These side channel attacks achieve specific targets, such as detecting the co-located programs running in VMs, destroying address randomization, enhancing rowhammer, and revealing secret information.

However, there are seldom work related to software resource-based covert channels, mainly including the page cache attack \cite{PCA19} and the container-based attack \cite{GaoSGKPW21}. In \cite{PCA19}, the Trojan sends data by accessing or not accessing the specific operating system page. Experimental results on Linux show that the channel's TR is up to 77.52 kb/s, and the average TR is 56.32 kb/s. In \cite{GaoSGKPW21} which is launched in multi-tenant cloud services, the Spy transmits information by reading dynamic identifiers and performance data of the memory-based pseudo-file (mainly procfs and sysfs). The unique dynamic identifiers-based (/proc/locks) channel provides 5.15 kb/s (8 locks) and 22.186 kb/s (32 locks) data TRs, with BER \textless~2\%. The performance data variation-based (/proc/meminfo) covert channel has a 13.6 b/s TR, less than the speed of unique dynamic identifiers-based covert channel. However, it can transmit data more reliable with 0.5\% BER.

Compared with the existing software resource-based covert channels \cite{PCA19,GaoSGKPW21}, the MES-Attacks have obvious advantages in the following aspects.


\textbf{Compared with page cache attacks \cite{PCA19}:} (1) Page cache attacks are more complex and requires page cache eviction, while the MES-Attacks can be achieved by calling the function directly; (2) Page cache attacks are implemented through a specific resource--operating system page cache, while the MES-Attacks can be realized through the resources available to all processes, such as variables and data structures. These resources appear in numerous scenarios and cannot be isolated, making them more difficult to defend.

\textbf{Compared with container-based attacks \cite{GaoSGKPW21}:} (1) Container technology provides a lightweight OS-level virtual hosting environment, which is applied to the development and deployment paradigm of multi-tier distributed applications. The main scenario is cloud services. Compared with container, MES-Attacks can be carried out in more working scenarios. Since the MESM that we utilize is one of the most basic technologies in OS. As long as the processor is concurrent, they are needed to ensure the smooth completion of the task. At the same time, we can create channels in cloud services by using files shared between containers as well. (2) \cite{GaoSGKPW21} leaks host information through a memory-based pseudo-file, which is changing all the time since the content displayed in the pseudo-file is internal kernel data. Especially in a cloud environment where noise exists, the contents might fluctuate tremendously.


\section{Conclusion}

To achieve better performance, modern operating systems are designed to execute concurrently with multi-processes, where MESM among processes is one of the basic mechanisms to ensure correct operation. However, the MESM can disclose resource usage among inter-constraint processes because  software resources in the OS are sharing. In this paper, we discover new covert channels based on mutual exclusion and synchronization, named MES-Attacks, and then construct a novel cooperation-based covert-channel mechanism. More importantly, due to the outstanding advantages of closed resources, detecting and mitigating MES-Attacks can be a daunting and lengthy task. We have implemented MES-Attacks in local, cross-sandbox and cross-VM scenarios, and they can achieve TRs of 13.105 kb/s, 12.383 kb/s, and 6.552 kb/s at less than 1\% error rate, respectively. 


\bibliographystyle{IEEEtranS}
\bibliography{main}

\end{document}